\documentclass[aps,prl,twocolumn,longbibliography]{revtex4-2}[12pt]

\usepackage[a4paper, total={7in, 10.3in}]{geometry}
\usepackage[utf8]{inputenc}
\usepackage[english]{babel}
\usepackage[font=small,labelfont=bf, justification=raggedright,singlelinecheck=false]{caption} 
\usepackage{amsfonts, amsmath, amsthm, amssymb} 
\usepackage{braket}
\usepackage{graphicx}
\usepackage{footnote}
\usepackage{tabularx}
\usepackage{array}
\graphicspath{ {./Figures/} }
\usepackage[plainpages=false,pagebackref=false,pdftex]{hyperref}

\hypersetup{colorlinks=true,breaklinks=true,linkcolor=black,citecolor=black, urlcolor=black}

\begin{document}
    
\title{Multi-scale calculation of light-induced structural changes in low-angle twisted bilayer WSe$_2$}

\author{Rafael R. Del Grande}
\email{rdelgrande@ucmerced.edu}
\author{David A. Strubbe}
\email{dstrubbe@ucmerced.edu}
\affiliation{Department of Physics, University of California, Merced, California, USA}

\date{\today}

\begin{abstract}

Exciton-phonon interactions in transition metal dichalcogenides (TMD) are strong and lead to phenomena such as coherent phonon generation. When stacked and twisted, their properties can be tuned by the twisting angle. In experiments with 1.1$^\circ$ twisted 2L WSe$_2$ [arXiv:2601.13620], an increase of 0.2 Å in the interlayer distance was observed when light was shone on this material, and here we explain the microscopic mechanism behind this.
Theoretical works to study such systems are limited because the moiré unit cell is too large. To overcome this, we combined classical force field relaxations with our implementation of \textit{ab initio} GW/Bethe-Salpeter excited state forces (ESF). From the relaxations we found that the low-angle twisting induced an in-plane strain field, the AB regions are large enough to be simulated as periodic AB stacked 2L WSe$_2$, and the interlayer force constant becomes softer in relation to the perfect AB stacking. From the \textit{ab initio} ESF we found that the in-plane strain increases the out of plane ESFs. Those two effects combined, the weakening of the interlayer force constant and strain dependence of the ESF, make light-induced changes in the interlayer distance of twisted 2L WSe$_2$ stronger than in the perfectly stacked case, in agreement with experimental observations.
Therefore, our results show that the exciton-phonon interactions can be tuned in twisted 2L TMDs and can be observed experimentally, which makes those materials excellent platforms to study light-induced changes in materials.

\end{abstract}

\maketitle

\textit{Introduction.} Transition metal dichalcogenides excitonic present strong excitonic properties that have been studied experimentally and theoretically. Exciton-phonon interactions in those materials have been probed through coherent phonon generation and photoluminescence experiments. Those layers can be stacked on top of each other, by Van der Waals interactions, and their electronic and optical properties depend on the the relative stacking. Those layers can also be twisted, which leads to more interesting phenomena \cite{Tran2020, Vitale2021, Regan2022}. 

Recently, a dark field tomography method was used to measure structural changes in 0.3° twisted bilayer WSe$_2$ \cite{Asuka2026arxiv}. In those experiments, the authors observed an increase of 0.2 \text{\AA} in the for the interlayer distance after shining a laser with energy equal to 2.4 eV. Subsequently, they observed that the interlayer distance returned to be what it was before the laser pulse in a time frame about 400ps, where they attributed this decay by two effects: a fast exciton-exciton annihilation and then a slow thermal decay. The used laser fluence is 330 $\rm{\mu J/cm^2}$, and assuming an optical absorption efficiency of 10\%, the estimated exciton concentration is about $\sim 8.6\times10^{13}$ excitons/cm$^2$ or 0.08 exciton per bilayer primitive cell (with two W and four Se atoms). Therefore one question arises: what is the mechanism behind the increase of the interlayer distance in twisted bilayer WSe$_2$? To answer this question we performed a multiscale study combining classical molecular dynamics of the twisted bilayer with \textit{ab initio} excited state forces calculations \cite{ESF_2025} of the AB-3R stacked bilayer. Our approximation is justified by the fact that small angle twisted bilayers are very similar to AB stacked bilayers as the AA and saddle point (SP) regions are much smaller than the AB region. Layer breathing mode (LBM) frequencies on twisted bilayers depend on the twisting angle, although for bilayer WSe$_2$ angles smaller than 3° have LBM frequencies independent of the twisting angle. Electron energy-loss spectroscopy experiments in twisted 2L WSe$_2$ show that energy of excitons A, B, C, and D for angles up to 28° vary about 0.1 eV \cite{WooPRB2023}, while photoluminescence measurements for twisting angles up to 35° of exciton A energy vary about 45 meV \cite{Barman2022ACSOmega}. Wu \textit{et al.} found that moiré depth potentials for A exciton of 109 and 215 meV for 1.36° and 3° twisted 2L WSe$_2$, respectively \cite{Wu2022LSA}.  Those energy variations are small compared to exciton energies (1.6 eV for exciton A), and are due to bandgap variations over the moiré supercell, and therefore we approximate excitonic effects of the twisted bilayer to be similar to the AB-3R stacked case. 

\textit{Computational details.} Our moiré supercell relaxations were done with the LAMMPS code \cite{LAMMPS}, using the Stillinger-Weber potential \cite{SW_PRB} for bonded interactions with parameters from \cite{Norouzzadeh_2017} and the Kolgomorov-Crespi potential \cite{Ouyang2018} for the interlayer interactions using parameters from  \cite{Naik2019}. We calculate the excited state forces (ESF) using our implementation \cite{ESF_2025} that combines excitonic effects from solutions of the Bethe-Salpeter Equation (BSE) with electron-phonon coefficients from DFPT. The ESF is given by the following equation
\begin{equation}
\begin{split}
    & F^S_{i,\alpha}  = - \sum_{\nu} \langle A | \nabla_\nu H^{\rm{BSE}}| A \rangle \hat{u}^\nu_{i,\alpha}\\
    & = - \sum_{\nu \mathbf{k}cvc'v'} A^*_{\mathbf{k}cv} A_{\mathbf{k}c'v'} (g^\nu_{c\mathbf{k},c'\mathbf{k}}\delta_{v,v'} - g^\nu_{v\mathbf{k},v'\mathbf{k}}\delta_{c,c'}) \hat{u}^\nu_{i,\alpha},
    \label{eq:excited_state_forces_our_method}
\end{split}
\end{equation}
where $i$ is one of the cartesian directions $x,y,z$, $\alpha$, is an atomic index, $\nu$ is the index of the displacement pattern $\boldsymbol{u}^\nu$ (i.e. a phonon mode), $g^\nu_{ik,jk} = \langle \psi_{ik} | \partial^\nu V| \psi_{jk} \rangle$ is the electron-phonon coefficient due to the collective atomic displacement $\nu$ and $A_{\mathbf{k}cv} = \langle A | \mathbf{k}cv  \rangle$ is the exciton coefficient in the basis of transitions $\mathbf{k}v \to \mathbf{k} c$ obtained by solving the BSE. DFT and DFPT calculations were performed with the Quantum ESPRESSO code \cite{Giannozzi2009, Giannozzi2017, Giannozzi2020} and GW/BSE calculations were performed with the BerkeleyGW code \cite{Deslippe2012, Rohlfing2000, Hybertsen1986}. We used ONCV scalar-relativistic LDA pseudopotentials with standard precision from Pseudodojo \cite{VansettenCompPhysCom2018, HamannPRB2013}. DFT calculations were performed with a cutoff energy of 80 Ry and the unitary cell vector perpendicular to the sheet planes was chosen to be  30 \text{\AA}. Before GW/BSE calculations, structures were relaxed with a force threshold of $1\times 10^{-3}$ eV/$\text{\AA}$ (1.4$\times$10$^{-4}$ Ry/bohr) and a cutoff energy of 160 Ry. We performed G$_0$W$_0$ calculations within the Generalized Plasmon Pole method \cite{Deslippe2012, Hybertsen1986}, with a cutoff energy to build the dielectric matrix equal to 25 Ry, and the stochastic pseudobands method \cite{AltmanPRL2024} with an accumulation window of 2\%, 2 stochastic pseudobands per energy subspace, and generating empty states with energy up to the DFT cutoff (80 Ry). Quasi particle (QP) energies and kernel matrix elements were calculated in a coarse $k$-grid 9$\times$9$\times$1 and interpolated into a fine $k$-grid 18$\times$18$\times$1 to build the BSE Hamiltonian. To improve the convergence with respect to $k$-grid in both GW and BSE calculations, we used the non-uniform sampling methods described in \cite{JornadaPRB2017}. 

\begin{figure}[h]
    \centering
    \includegraphics[width=\linewidth]{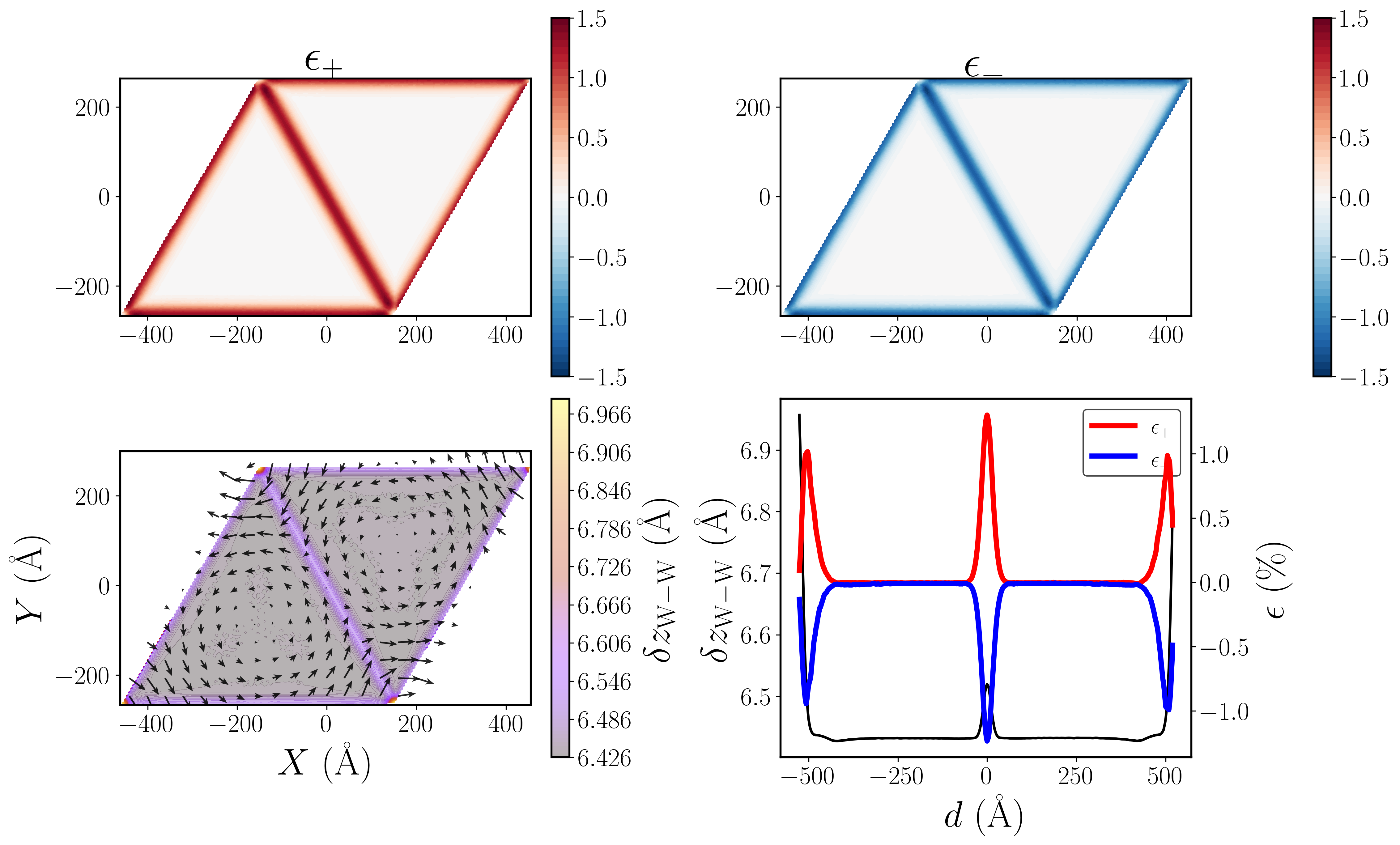}
    \caption{(a) and (b) are the eigenvalues $\epsilon_+$ and $\epsilon_-$ of the strain tensor over the moiré supercell. (c) atoms displacement field due to relaxation. (d) Strain over the moiré supercell diagonal. In the AB regions the strain is zero, although it is anisotropic in the AA and SP regions.}
    \label{fig:twisted_bilayer_relaxation}
\end{figure}

\textit{Results and discussion.} First we focus on the relaxations of the moiré patterns, as after twisting the atomic reconstruction plays a key rule in the electronic and optical properties of twisted layered materials \cite{CulchacNanoscale2020, CulchacPRB2025}. The reconstruction process creates a local strain in those materials that is dependent on the twist angle \cite{Halbertal2022, Rodriguez2023ACSNano}.
The relaxed structure presents large triangular regions with AB stacking and small regions with AA stacking, as shown in Figure \ref{fig:twisted_bilayer_relaxation}. From the relaxation process we obtained the displacement fields for atoms centered at $\mathbf{r}$ is given by $\mathbf{u}(\mathbf{r)}$. On figure \ref{fig:twisted_bilayer_relaxation} we can see that the displacement field rotates around the AA area, where the atomic reconstruction tries to avoid AA stacking and favors AB stacked areas. After computing the displacement fields one needs to perform numerical derivatives to calculate strain. To avoid numerical instabilities when interpolating the displacements $\mathbf{u}$ we used an alternative approach. For a given layer the set of W atoms form a triangular lattice, so we use that the fact that three non-collinear points define a plane. Each component $\alpha=x,y$ of the displacement field $\mathbf{u}$ at the center $\mathbf{\bar{r}}=(\bar{x}, \bar{y})$ of a triangle with edges $\mathbf{r}_1, \mathbf{r}_2$ and $\mathbf{r}_3$ obeys a plane equation on the form $\mathbf{u}_\alpha (x,y) \approx \partial_x \mathbf{u}_\alpha (x-\bar{x}) + \partial_y \mathbf{u}_\alpha (y-\bar{y}) + \mathbf{u}^0_\alpha$ and the spatial derivatives of the displacement field can be obtained by

\begin{equation}
    \begin{bmatrix}
    \frac{\partial \mathbf{u}_\alpha}{\partial x} \\
    \frac{\partial \mathbf{u_\alpha}}{\partial y} \\
    \mathbf{u}_{\alpha}^0
    \end{bmatrix}
    = {
    \begin{bmatrix}
        x_1-\bar{x} & y_1-\bar{y} & 1 \\
        x_2-\bar{x} & y_2-\bar{y} & 1 \\
        x_3-\bar{x} & y_3-\bar{y} & 1 \\
    \end{bmatrix}}^{-1}
    \begin{bmatrix}
        \mathbf{u}_{\alpha, 1} \\
        \mathbf{u}_{\alpha, 2} \\
        \mathbf{u}_{\alpha, 3}
    \end{bmatrix}
\end{equation}
and then the strains are computed using the relations $\epsilon_{xx} = \partial_x \mathbf{u}_x$, $\epsilon_{yy} = \partial_y \mathbf{u}_y$ and $\epsilon_{xy} = 1/2( \partial_x \mathbf{u}_y + \partial_y \mathbf{u}_x)$. This procedure is numerically stable and does not depend on choices of grid sizes for numerical interpolation. For each W-triangle we then find a $2\times 2$ strain matrix, which eigenvalues are $\epsilon_{\pm}$. We found that the strain is anisotropic in the saddle point (SP) and AA regions reaching $\pm 1\%$, and $0\%$ in the AB region, as shown in Figure \ref{fig:twisted_bilayer_relaxation}. For larger angle twisted bilayers the AB region in the moiré unit cell becomes smaller and it will have some anisotropic strain. 



\begin{figure}[!h]
    \centering
    \includegraphics[width=\linewidth]{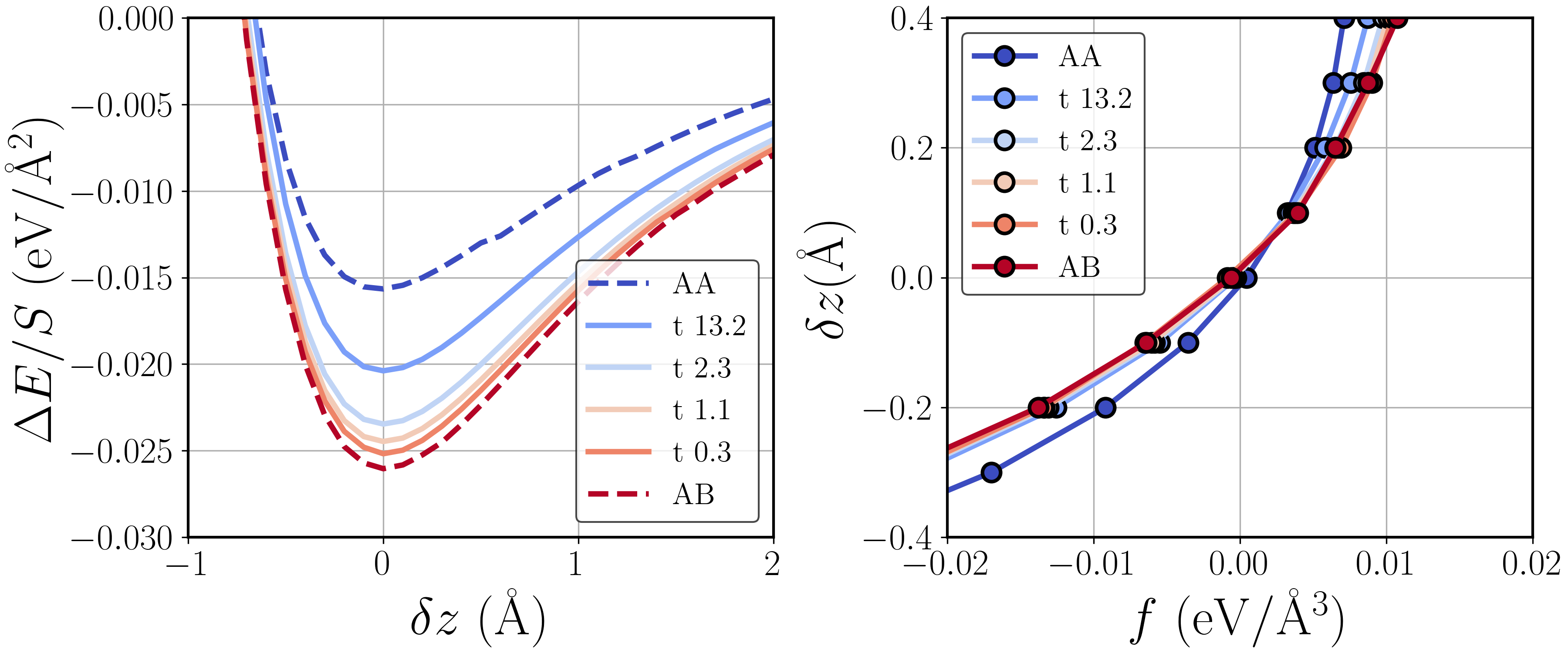}
    \caption{Left: Total cohesive energy density for AA, AB, and twisted bilayer WSe$_2$ for some angles. Small angles are close to AB stacked case while larger angle cases are close to AA stacking. Right: Interlayer distance change as function of the repulsive force density applied on the bilayer. Forces density are calculated by finite differences using the data from left panel.}
    \label{fig:E_vs_d_2L}
\end{figure}

We performed MD calculations to study how the cohesive energy density changes when the bilayers move rigidly in relation to each others for the AB and AA stackings and some twisted cases, and those results are summarized in figure \ref{fig:E_vs_d_2L}. Smaller angles energy density curves are closer to AB stacking, while smaller angles are closer to AA stacking. This agrees with the fact that in smaller angle twisted bilayers the AB region becomes larger in the moiré unit cell. The atomic reconstruction makes the frequency of layer breathing modes for angles smaller than 3° to be closer to AB stacked frequencies \cite{Lin2021AdvMat}.
Then we calculated the force density by numerical differentiation $f=-\partial (\Delta E/S)/\partial z$ and those results are shown in figure \ref{fig:E_vs_d_2L} (b). We can see that those forces are asymmetrical with respect to the displacements from equilibrium $\delta z$, and AA stacked bilayers interlayer force constants are weaker. We estimate that the necessary force density to induce 0.2 \textbf{\AA} interlayer separation as in \cite{Asuka2026arxiv} is about $6 \times 10^{-3}$ \text{eV/\AA}. 

\begin{figure}[!h]
    \centering
    \includegraphics[width=\linewidth]{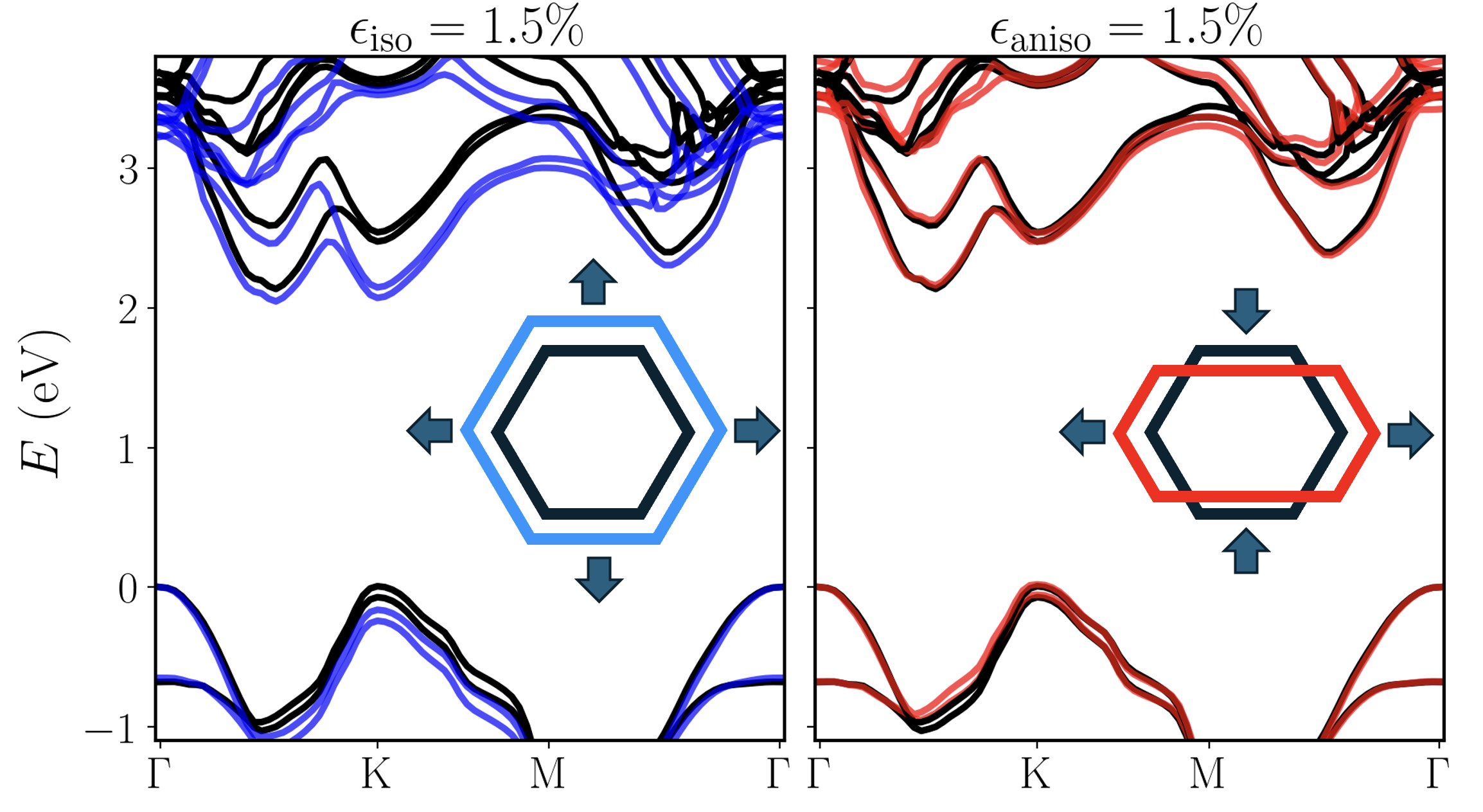}
    \caption{Bandstructure for bilayer WSe$_2$ under isotropic (left) and anisotropic (right) strain}
    \label{fig:bands_iso_vs_aniso_strain}
\end{figure}

Now we move to \textit{ab initio} results. First we study the effect of anisotropic strain on the electronic properties of bilayer WSe$_2$. In figure \ref{fig:bands_iso_vs_aniso_strain} we show that isotropic strain induces substantial change in WSe$_2$. For isotropic strains larger than 2\% WSe$_2$ goes from a direct to an indirect semiconductor \cite{Tang2020}. On the other side, anisotropic strain induces small changes in the band structure. This can be explained with the theory of the deformation potential, where changes in the electronic properties are typically linear with the trace of the strain tensor while for anisotropic strain this dependence is quadratic, which for small strains is negligible.

As the AB region is sufficiently large, we approximate the optical properties of this material to be the same as the properties of the a 3R non twisted bilayer. In twisted bilayer WSe$_2$ exciton energies for A, B, C and D excitons vary only about $\sim 0.1$ eV \cite{WooPRB2023, Barman2022ACSOmega}, even though there is the formation of flat bands for small angles \cite{Vitale2021}. In twisted bilayer (T2L) MoS$_2$, when the twisting angles goes from 30 to close to zero degrees the A and B exciton energies remain unaltered, while the finite momentum exciton energy decreases \cite{Liao2020}, so intralayer transitions remain unaltered. Here we are interested in exciton with no momentum, namely the excitons A, B and C \cite{McCrearyNanoscale2022}, which may remain unaltered in low-angle twisted bilayer WSe$_2$, justifying our approximation. WS$_2$/WSe$_2$ heterostructures near 1 degree have the intralayer exciton with energy approximately constant  \cite{Maity2025npj2d}. 

\begin{figure}[h]
    \centering
    \includegraphics[width=\linewidth]{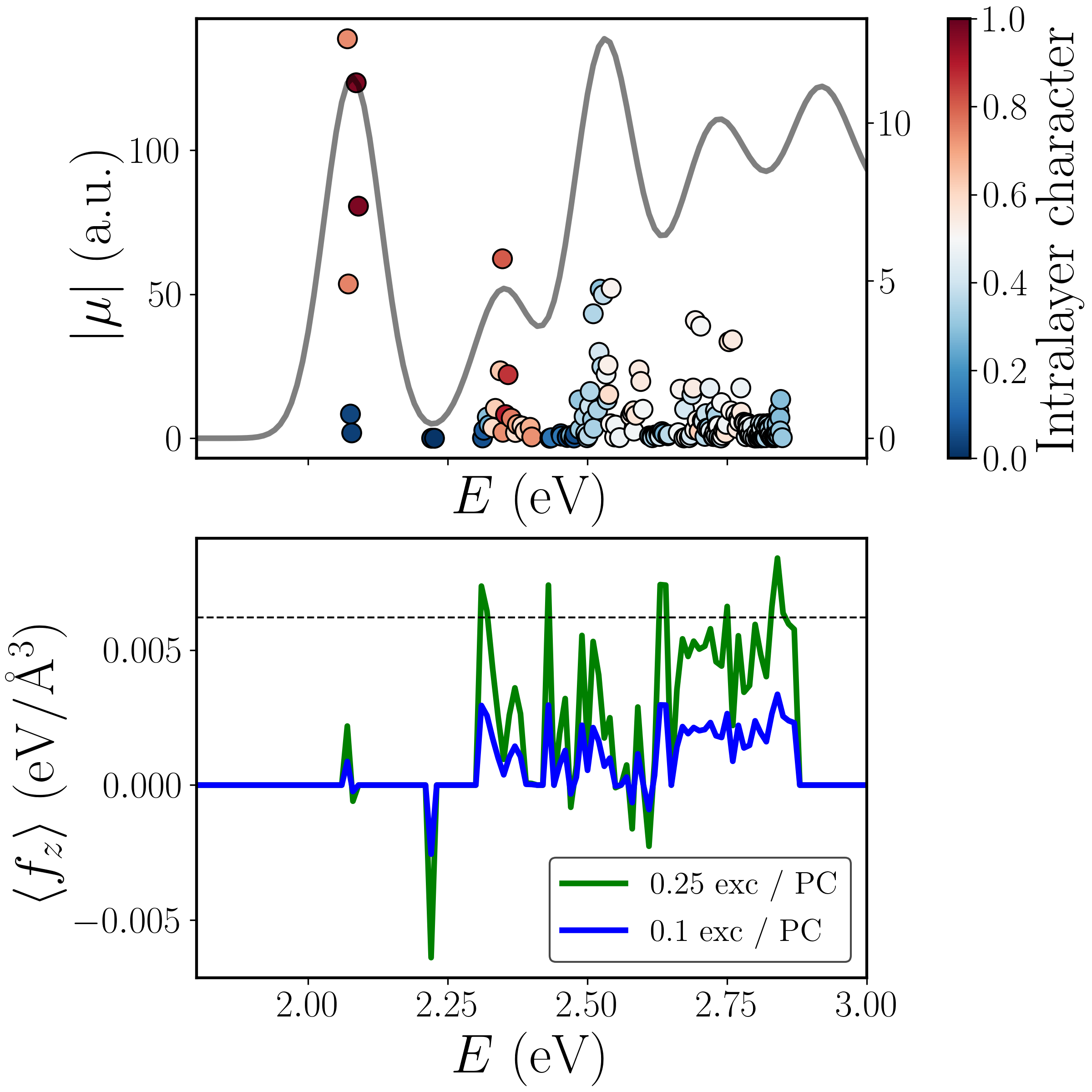}
    \caption{Upper panel: Solid black line: optical absorption, while circles are the exciton transition dipole moment. Color grading goes from blue (interlayer) to red (intralayer) using the intralayer character defined by equation \ref{eq:intralayer_character}. Lower panel: averaged excited state force on layers of bilayer WSe$_2$, (see equation \ref{eq:dip_mom_weighted_average_esf}). Positive (negative) forces are repulsive (attractive).}
    \label{fig:dip_mom_weighted_average_esf}.
\end{figure}

Then we compute the excited state forces (eq. \ref{eq:excited_state_forces_our_method}) \cite{ESF_2025}. We studied ESF projected out-of-plane directions. Each absorption peak is composed by a diversity of excitons, therefore we calculate averages of the excited state forces, using the transition dipole moments $\mu^i = \langle 0|\mathbf{r}|S_i \rangle$, using the following equation
\begin{equation}
    \langle f_z\rangle (E)= \frac{\sum_i  f_z^i |\mu^i|^2}{\sum_i  |\mu^i|^2} \rm{ \ for \  } |E-\Omega_i| \leq \delta E
    \label{eq:dip_mom_weighted_average_esf}
\end{equation}
with $\delta E = 0.05$ eV. 
In figure \ref{fig:dip_mom_weighted_average_esf} we show the average of the ESF for exciton concentrations of 0.1 and 0.25 exciton per primitive cell (PC). For a given exciton concentration $x$ the total excited energy for a calculation will be $x \Omega$ per unit cell and the ESF will be $-x \nabla \Omega$ per unit cell. In \cite{Asuka2026arxiv} an exciton concentration $\sim$ 0.1 exciton per primitive cell ($\sim 1.1 \times 10^{14}$ exc/cm$^2$) is responsible for an interlayer distance increase of 0.2 \text{\AA}. We see that our results get close to the necessary ESF to induce when we use an exciton concentration five times from what authors of \cite{Asuka2026arxiv} estimated. Besides this factor discrepancy we have found that the majority of the excited state forces are repulsive ($\langle f_z\rangle >0$) instead of attractive, specially for excitations close to the excitation energy used in \cite{Asuka2026arxiv} of 2.4 eV, close to exciton C. The majority of excitons that compose peaks A, B and C are involve transitions from the first two valence bands to the first conduction bands and happen at the K point. In a tight-binding model for transition metal dichalcogenides the first conduction and valence bands close to the K point form bonding and anti-bonding pairs, and in an analogy of H$_2$ molecule, where the distance between the layers increases the energy separation between the bonding and anti-bonding energy levels so therefore the excited state force is repulsive. 

To classify excitons as interlayer or intralayer we use the out of plane exciton density defined by
\begin{equation}
    \rho(z_e, z_h) = \int |\Psi_{ex} (\mathbf{r}_e, \mathbf{r}_h) |^2 dx_e dy_e dx_h dx_e
\end{equation}
where the exciton wavefunction is $\Psi_{ex} = \sum_{\mathbf{k}cv} A_{\mathbf{k}cv}\psi^*_{\mathbf{k}c}(\mathbf{r}_e) \psi_{\mathbf{k}v}(\mathbf{r}_h)$, with $\psi_{\mathbf{k}n}$ being the single particle states obtained from DFT and $\mathbf{r}_{e(h)}$ being the position vector of the electron (hole). The coordinate $z$ is chosen to be perpendicular to the bilayer WSe$_2$. We classify excitons as interlayer or intralayer using our definition of intralayer character as
\begin{equation}
    x=\int_{|z_e-z_h|<d/2} \rho(z_e,z_h)dz_e dz_h
    \label{eq:intralayer_character}
\end{equation}
which is a number between 0 and 1, and values close to 1 (0) represent an intralayer (interlayer) exciton. Using this classification, we can see that excitons that compose peaks A and B are more intralayer excitons, while peak C is composed by interlayer excitons. We found no clear correlation between the exciton character with the intensity or if the correspondent excited state force is repulsive or attractive.

\textit{Conclusions.} Our molecular dynamics indicate that mechanical properties of low angle twisted bilayers are close to AB stacked and to induce an increase of 0.2 \text{\AA} in the interlayer distance the necessary force density is about $6\times 10^{-3}$ eV/\text{\AA$^3$}. In the AB region there is no strain, although in the AA and SP regions the strain is anisotropic and the trace of the strain tensor is zero. Using this information we see in our \textit{ab initio} calculations that the effect of anisotropic strain is negligible in the electronic and optical properties of bilayer WSe$_2$. The excited state forces projected in the out of plane direction for excitons with energy in the range from 2.0 to 2.8 eV are in general repulsive and strong enough to induce changes in the interlayer distance (of the order of $\sim 0.1$ \text{\AA}) detectable by experiments. In plane shear forces are also sufficiently to induce distortions such twist. Excited state forces parallel to covalent bonds are stronger than interlayer forces, but are not strong enough to induce substantial displacements as the covalent bonds W-Se are much stronger than the Van der Waals interaction between the layers. Therefore the exciton-phonon interactions in multilayer systems can be probed by relative twist between the layers and changes in the interlayer distance.  


\textit{Acknowledgments.} We thank Asuka Nakamura for fruitful discussions about his experiments. R.R.D.G. and D.A.S. were supported by the U.S. National Science Foundation under Grant No. DMR-2144317. Computational resources were provided by the National Energy Research Scientific Computing Center (NERSC), a U.S. Department of Energy Office of Science User Facility operated under Contract No. DE-AC02-05CH11231 using NERSC awards DDR-ERCAP0031916 and DDR-ERCAP0038053; the Texas Advanced Computing Center (TACC) at The University of Texas at Austin (http://www.tacc.utexas.edu); and the Pinnacles cluster (NSF MRI, \# 2019144) at the Cyberinfrastructure and Research Technologies (CIRT) at the University of California, Merced.


\bibliography{refs}



\end{document}